# An Overview: Extensible Markup Language Technology

Rami Alnaqeib, Fahad H.Alshammari, M.A.Zaidan, A.A.Zaidan, B.B.Zaidan, Zubaidah M.Hazza

**Abstract —** XML stands for the Extensible Markup Language. It is a markup language for documents, Nowadays XML is a tool to develop and likely to become a much more common tool for sharing data and store. XML can communicate structured information to other users. In other words, if a group of users agree to implement the same kinds of tags to describe a certain kind of information, XML applications can assist these users in communicating their information in an more robust and efficient manner. XML can make it easier to exchange information between cooperating entities. In this paper we will present the XML technique by fourth factors Strength of XML, XML Parser, XML Goals and Types of XML Parsers.

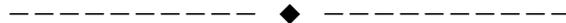

## 1. INTRODUCTION

In 1986 the Standard Generalized Markup Language (SGML) became an international standard for defining descriptions of the structure and content of different types of electronic documents. SGML, the "mother tongue" of HTML and XML, is used for describing thousands of different document types in many fields of human activity, from transcription of ancient Sumerian tablets to the technical documentation for steel bombers, and from patient's clinical records to musical notations [1].

SGML has withstood the test of time. Its popularity is rapidly increasing among organizations with large amounts of document data to create, manage and distribute as in the Defence, Aerospace, Semiconductors and Publishing industries. However, various barriers exist to delivering SGML over the Web. These barriers include the lack of widely supported style sheets, complex and unstable software because of SGML's broad and powerful options, and obstacles to the interchange of SGML data because of varying levels of SGML compliance among SGML software packages [2].

These difficulties have condemned SGML to being a successful niche technique rather than a mainstream tool. Indeed some cynics have renamed SGML in 'Sounds Good Maybe Later' [3].

HTML (Hyper Text Markup Language) is a subset of SGML, the most frequently used document type in the Web. It defines a single, fixed type of document with markup that lets you describe a common class of a simple office style report, with headings, paragraphs, lists, illustrations, etc., and some provision for hypertext and multimedia [4].

HTML was defined to allow the transfer, display and linking of documents over the internet and is the key enabling technology for the WWW. Prior to the emerging of the internet, it was unusual in the word of computing to hear the word "page" used to describe elements of data.

HTML web pages have amazing similarities with paper in their role of information publishing [5].

HTML was designed as a markup language with simple structures, strong emphasis on formatting and weak for encoding content. It was not designed to encode structure and semantics needed for complex applications. Because of the lack of SGML support in mainstream Web browsers, most applications that deliver SGML information over the Web convert the SGML to HTML. This down-translation removes much of the intelligence of the original SGML information. That lost intelligence virtually eliminates the flexibility of information and poses a significant barrier upon reuse, interchange and automation [6].

For this reason, XML (Extensible Markup Language) was developed by the XML working group (known as the SGML Editorial Review Board) formed under the auspices of the W3C in 1996 (W3C is the World Wide Web Consortium. The W3C is dedicated (in part) to encouraging the development of open Web standards, such as the HTML and XML document markup languages, to promote interoperability and assist the Web in achieving its potential.) [7].

XML is a highly functional subset of SGML. The purpose of XML is to specify an SGML subset that works very well for delivering SGML information over the Web. When the mainstream Web browsers support XML, it is believed that it's going to be very easy to publish SGML information on the Web. It is actually misnamed because XML is not a single Markup Language. It is a meta language that allows users to design their own markup language [1].

XML parser is a software module that reads documents and provides access to their content. It generates a structured tree to return the results to the browser. The parser is similar to a processor that determines the structure and properties of the data. It can read an XML document to create an output and to generate a display form consequently [2].

Yanming proposed a new method for designing and implementing a manual XML parser named BNFParser [17]. It is based on the mechanism of XML document matching against a BNF tree that builts on XML formal grammar represented by Extended Backus-Naur Form (EBNF) notation. Compared with generic XML parsers, e.g. Top-down Xerces/Libxml parser and Bottom-up YACC–assisted Parser, BNFParser is designed to be used under the circumstance of storage limitation and memory restriction, e.g. embedded devices [8].

The advantage of BNFParser is that it is far more flexible than to be tailored and upgraded through only updating plain text BNF syntax rules instead of reediting and recompiling source code. Furthermore, the architectures of the generic XML parsers and the performance evaluation were investigated. Experiment showed that applicable scenario of BNFParser is parsing an XML document with size less than 100 KB [17].

Zhang and Engelen presented TDX, a table-driven XML parser [18]. TDX combines parsing and validation into one pass to increase the performance of XML-based applications, such as Web services. The TDX approach is based on the observation that context-free grammars can be automatically derived from XML schema [10].

They developed a parser construction tool to automatically construct TDX grammar productions from a schema. Grammar tokens are defined by the specific schema element names, attribute names, and text. Because most of the structural constraints in XML schema are cast as grammar rules, parsing and validation of XML instances are efficiently implemented. The results showed that TDX is several times faster than DOM or SAX parsing with enabled validation [11].

A number of researchers understood that a poor performance in parsing / loading XML documents might constitute a serious obstacle to effectively adopt XML-based solutions in applications like a web services and e-business systems; therefore, several researches are trying to address this problem by improving the parsing phase, e.g., by adopting condensed or binary representations of XML documents [19]. Researchers are working on this topic and some results are arising: for example; Takes et al  [20] exploited efficient data structures that speed-up XML parsing when parsed XML documents belong to a limited family of documents (such as in the context of web services)[12],[13].

In another approach, researchers are investigating the so-called Schema-Specific Parsing [21]: based on a given XML Schema specification, a parser is generated to only recognize XML documents compliant with the source XML Schema specification. This approach is good in contexts where the XML document family to process, is known in advance.

A third approach is the development of hardware accelerators [22],[23], specifically developed for very fast XML document parsing. This solution is suitable for embedding XML processing within devices, but not for general purpose computers [17],[18],[19].

A different way to reduce XML parsing time is to change XML: this is the idea behind no less than 19 proposals for binary representation of XML documents [23], known as Binary XML. The idea is that a binary representation is processed faster than a textual one. A specific W3C Working Group, named XBC XML Binary Characterization Working Group, is now working on the problem [14],[15],[16].

## 2. XML Technology

XML is a text-based markup language that is fast becoming the standard for data interchange on the Web. As with HTML, data is identified by the usage of tags (identifiers enclosed in angle brackets, like this: <…>). Collectively, the tags are known as "markup".

Unlike HTML, XML tags identify the data, rather than specifying how to display it. While an HTML tag says something like "display this data in bold font" (<b>…</b>), an XML tag acts like a field name in the program. It puts a label on a piece of data that identifies it (for example: < TITLE >…</ TITLE >).

Since identifying the data gives some sense of what it means (how to interpret it, what you should do with it), XML is sometimes described as a mechanism for specifying the semantics (meaning) of the data [3].

```
<?xml version="1.0"?>
<!--File Name: Inventory.xml -->
<INVENTORY>
      <BOOK>
              <TITLE>The Adventures of Huckleberry Finn</TITLE>
              <AUTHOR>Mark Twin</AUTHOR>
              <BINDING>Mass Market Paperback</BINDING>
              <PAGES>298</PAGES>
              <PRICES>$5.49</PRICES>
      </BOOK>
      <BOOK>
              <TITLE>Leaves of Grass</TITLE>
              <AUTHOR>Walt Whiteman</AUTHOR>
              <BINDING>Hardcover</BINDING>
              <PAGES>462</PAGES>
              <PRICES>$7.75</PRICES>
      </BOOK>
      <BOOK>
              <TITLE>The Legend of Sleepy Hollow</TITLE>
              <AUTHOR>Washington Irving</AUTHOR>
              <BINDING>Mass Market Paperback</BINDING>
              <PAGES>98</PAGES>
              <PRICES>$2.95</PRICES>
      </BOOK>
</INVENTORY>
```

Fig 1.  Example of XML File

The above file describes an inventory list. INVENTORY represents the root element. The document contains three nested BOOK elements. Every BOOK element has five sub elements: TITLE, AUTHOR, BINDING, PAGES and PRICES [4].

## 2.1 Strength of XML

Some features of XML that make it well-suited for data transfer are:

- It is being simultaneously human and machine-readable format;
- It supports Unicode, allowing almost any information in any written human language to be communicated;
- It can represent the most general computer science data structures: records, lists and trees;
- Its self-documenting format describes structure and field names as well as specific values;
- The strict syntax and parsing requirements make the necessary parsing algorithms extremely simple, efficient and consistent.

XML is also heavily used as a format for document storage and processing, both online and offline, and it offers several benefits:

- Its robust, logically-verifiable format is based on international standards;

- The hierarchical structure is suitable for most (but not all) types of documents;
- It is manifested as plain text files, unencumbered by licenses or restrictions;
- It is platform-independent, thus relatively immune to changes in technology;
- Its predecessor, SGML, has been in use since 1986, so there is an extensive experience and software availability [5].

## 2.2 XML Goals

The Goals set for the XML by the W3C Working Group are as follows:

- The XML will be straightforwardly usable over the Internet.
- It will support a wide variety of applications.
- It will be compatible with SGML.
- It will be easy to write programs which process XML documents.
- The number of optional features in XML is to be kept to the absolute minimum, ideally zero.
- XML documents should be human-legible and reasonably clear.
- The XML design should be prepared quickly.
- The design of XML will be formal and concise.
- XML documents will be easy to create.
- Terseness in XML markup is of minimal importance [6].

## 2.3 XML Parser

A parser is a computer program or a component of a program that analyses the grammatical structure of an input with respect to a given formal grammar in a process known as parsing. Typically, a parser transforms some input text into a data structure that can be processed easily, e.g. for semantic checking, code generation or to help understanding the input. Such data structure usually captures the implied hierarchy of the input and forms a tree or even a graph [7].

XML parser can be defined as a software (or a class in JAVA) that reads XML, checks for its conformance to standard, and/or also validates it. For any client or server application that needs to process XML, XML parser is definitely a must. XML would not be able to perform as desired before it has been parsed, despite its increasing popularity among developers and also commercial users. Therefore, the importance of XML parser has become significant in this matter.

## 2.4 Types of XML Parsers

Currently there are a lot of XML parsers, and most of them evolve, improve and become sophisticated. Though all the parsers serve the same purpose, they vary in terms of specification, performance, reliability and also conformance to standard. If a wrong choice has been made, it is highly possible to lead to excessive hardware requirement, poor system performance; productivity degradation as well as stability issues and the outcome would be disastrous. Based on the W3C recommendation, every XML document needs to be well-formed and valid. An XML document is well-formed if it fulfils three main criteria, namely: there are no overlapping tags, each opening tag must have a corresponding closing tag and the document must be in accordance with the specification recommendation [8].Otherwise, it would be considered as non well-formed [9].

XML parsers are classified along two independent dimensions: which is validating versus non-validating, in terms of functionality [10]; and stream-based versus tree based, in terms of Application Programming Interfaces (API) [11].A validating parser can use a Document Type Definition (DTD) or an XML Schema Definition (XSD) to verify that a document is properly constructed according to the rules of XML application [12]. As for a non-validating parser, it only requires the document to be well formed. Non-validating is relatively simpler compared to validating parser. Aparser can read the XML document components via (APIs) in two approaches. For a stream-based approach, it reads through the document and signals the application every time a new component appears. This is known as event-based approach. As for tree-based approach, it reads the entire document and gives the application a tree structure corresponding to the element structure of the document [12]. There are a numbers of APIs for XML, for example Document Object Model (DOM), Simple API for XML (SAX), Java-based Document Object Model (JDOM), Java API for XML Processing (JAXP), XMLPull and The Streaming API for XML (StAX).SAX, StAX, and XMLPull are event-based API approach while DOM, JDOM, electric XML, and DOM4j are categorized as tree-based API. Most of the major XML parsers support both SAX and DOM. However, there are a few parsers that only support SAX, and at least a couple that only support their own proprietary API like ElectricXML or XMLPull parser.

Table 1:

Illustrates a Brief Comparison Between XML Parser's APIs, with Respect to their Characteristics

| Parser APIs | Advantages | Disadvantages |
|---|---|---|
| DOM | - Rich set of APIs<br>- Easy navigation<br>- Entire tree loaded into memory, random access to XML document | - XML document must be parsed at one time<br>- Expensive to load entire tree into memory<br>- Generic DOM node not ideal for object-type binding |
| SAX | - Entire document is not loaded into memory, resulting in low memory consumption<br>- Allows registration of multiple Content Handlers | - No built-in document navigation support<br>- No random access to XML document<br>- No support for modifying XML in place<br>- No support for namespace scoping |
| StAX / XMLPull | - Contains two parsing models, for ease of performance<br>- Application controls parsing, easily supporting multiple inputs<br>- Powerful filtering capabilities provide efficient data retrieval | - No built-in document navigation support<br>- No random access to XML document<br>- No support for modifying XML in place<br>- Still in an immature state |
| ElectricX ML | - Light weighted<br>- Fast in performance | - No support for validating<br>- Still in immature state |

Elliotte has conducted a SAX conformance test using W3C XML conformance Test Suite on a number of parsers, where Xerces is the most conformant parser to SAX standard [13]. Another conformance testing with respect to XML standard using OASIS XML 1.0 Test Suites is carried out and result shows that Xerces has emerged to be the most conformant among other parsers [14]. Mohseni conducted a performance test in which Micorosft XML (MSXML) rivals other parser having the shortest load time [15]. Yet, another DOM based parser benchmark test performed by Sosnoski using XMLBench, which was tested on various criteria on DOM parsers didn't prove to be the best option. Xerces seems to outperform other parsers [11].

To encapsulate the above, the Xerces parser has emerged to be the best parser, as it provides support for a lot of XML and API standards. Besides, it is the most conformant and has best performance that rivals other parsers on test. Furthermore, it was also voted as the best XML parser of the year 2002 by XML-Journal/Web Services Journal Readers' Choice Awards [16].

## 3. Conclusion

In this paper we presetaed the XML technique by fourth factors Strength of XML, XML Parser, XML Goals and Types of XML Parsers as the last fator we present the table illustrates a Brief Comparison Between XML Parser's APIs, with Respect to their Characteristics.

### ACKNOWLEDGMENT

This research has been funded in part from multimedia University; the author would like to acknowledge the entire worker in this project, and the people who support in any way, the author would like to thanks his friends who has support in many ways